\documentclass[aps,prb,twocolumn,floatfix]{revtex4}
\usepackage{graphicx,epsfig,array,color}
\usepackage{color,graphicx,amssymb,amsmath}
\usepackage[utf8]{inputenc}
\begin{document}
\title{Orbital Selective Directional Conductor in the 
Two Orbital Hubbard Model}
\author{Anamitra Mukherjee$^{1,3}$}
\author{Niravkumar D. Patel$^1$}
\author{Adriana Moreo$^{1,2}$}
\author{Elbio Dagotto$^{1,2}$}
\affiliation{$^1$Department of Physics and Astronomy, The University of Tennessee, Knoxville, Tennessee 37996, USA}
\affiliation{$^2$Materials Science and Technology Division, Oak Ridge National Laboratory, Oak Ridge, Tennessee 37831, USA}
\affiliation{$^3$School of Physical Sciences, National Institute of Science Education and Research, Jatni 752050, India}

\begin{abstract}
Employing a recently developed many-body technique that allows for the incorporation
of thermal effects, the rich phase diagram of a 
two dimensional two orbital (degenerate $d_{xz}$ and $d_{yz}$) Hubbard model 
is presented varying temperature and the repulsion $U$.
Our main result is the finding at intermediate $U$ of a novel antiferromagnetic  
orbital selective state where an effective
dimensional reduction renders one direction insulating and the other metallic.
Possible realizations of this state are discussed. In addition we also study nematicity
above the N\'eel temperature. After a careful finite-size scaling analysis the nematicity temperature
window appears to survive in the bulk limit although it is very narrow.
\end{abstract}

\maketitle

\section{Introduction}

Several important materials, such as iron-based high critical
temperature superconductors (FeSC),\cite{ironrefs,peter,dai,RMP} 
nickelates,\cite{nirefs} cobaltites,\cite{corefs} 
manganites,\cite{mang-book,salamon} and many others, have several active orbitals.
If electronic correlation effects are important, then multiorbital 
Hubbard models must be employed for their analysis. However, 
these models are difficult to study
in two (2D) or three (3D) dimensions due to their complexity. 
While at particular electronic 
densities, such as integer fillings, zero temperature
mean field (MF)  approximations are often reliable, the phase diagrams of
multiorbital Hubbard models 
varying {\it temperature} $T$ are basically unknown. This is because
thermal mean-field approximations provide qualitatively incorrect results at 
robust Hubbard repulsion $U$ and intermediate temperature since
they cannot generate local moments with short-range magnetic order because of the absence of spatial fluctuations. 
Moreover, ``sign problems'' complicate the application of
quantum Monte Carlo (MC) when several orbitals are active.\cite{signproblems}
Since not only $U$ but also the Hund coupling $J$ 
are important, it is imperative to apply alternative computational
tools, even if crude, to study multiorbital models at intermediate temperatures because 
the potential for new states in these systems is considerable.

Recently,\cite{mfmc-rec-prb} the ``Monte Carlo-Mean Field'' (MC-MF) technique 
that mixes the MC and MF
approximations was tested using the half-filled 
one-orbital Hubbard model. The method captured all qualitative
features of this model known from quantum Monte Carlo, 
including the non-monotonic behavior of the N\'eel temperature 
$T_N$ with increasing $U$, and it was
even quantitatively accurate 
with errors of only $\sim$20\% in $T_N$.\cite{mfmc-rec-prb} 
This method was applied before to the BEC-BCS crossover in the 
cold atom context and to other problems.\cite{majumdar-many}
Early studies
within the spin-fermion model context showed that this type of techniques
are also reliable in studies of superconductors.\cite{mayr1,alvarez1,mayr2}

The new method relies on 
the Hubbard-Stratonovich decomposition of the interacting problem 
via auxiliary fields (AuxF)~\cite{majumdar-many,mayr1,alvarez1,mayr2} 
(in the Hartree channel in our study, but could be in 
other channels as well). Neglecting the AuxF's
imaginary-time dependence but retaining their spatial fluctuations 
leads to a Hamiltonian with quantum fermions coupled to classical degrees of freedom, 
similarly as in double-exchange models
for manganites.\cite{mang-book}
Classical Monte Carlo is used for the AuxF at any temperature, while 
the fermionic sector is treated via
the traveling cluster approximation (TCA) that allows 
access to large lattices.\cite{tca-1,tca-2}

In this publication, the MC-MF technique is applied for the first time 
to a 2D two-orbital Hubbard model, varying temperature $T$ and repulsion $U$. 
The unveiled phase diagram is rich, including 
a narrow nematic phase above $T_N$.\cite{Chu,nematicrefs} Even more importantly, here we report 
an unexpected novel regime, dubbed Orbital Selective Directional Conductor (OSDC), 
where a remarkable anisotropy in transport is observed, with 
one direction insulating and the other conducting, leading to a dimensional
reduction from 2D to 1D. This dimensional reduction is different 
from that in Tl$_2$Ru$_2$O$_7$ and BaCuSi$_2$O$_6$  
because they are insulating in all directions,~\cite{TRO,BCSO}
and different from layered Sr$_3$Ru$_2$O$_7$  because it requires 
a high magnetic field.~\cite{andy}
Our results are also different from layered oxides 
that are metallic in-plane but insulating out-of-plane,\cite{layeredTMO} 
because their crystal structure already establishes an asymmetry.
On the contrary, our two-dimensional model is fully symmetric 
between the $x$ and $y$ directions but {\it spontaneously}
becomes insulating in one direction and metallic in 
the other, without the help of the lattice, magnetic fields, or impurities.

\section{Model and Method} 

Although five orbitals are needed for a faithful electronic 
description of iron superconductors, below
for simplicity we will focus on the two most important orbitals 
$d_{xz}$ and $d_{yz}$.\cite{raghu,moreo,fesc-2orb-utk-1,fesc-2orb-utk-2}
The two-orbital Hubbard model studied here is defined as:

\vspace{-0.2cm}
\begin{equation}
\begin{aligned}
H = \sum_{\langle i,j \rangle,\alpha,\beta,\sigma} 
T^{i,j}_{\alpha,\beta} d^\dagger_{i,\alpha,\sigma} d^{\phantom\dagger}_{j,\beta,\sigma}
+ U \sum_{i,\alpha} n_{i,\alpha,\uparrow}n_{i,\alpha,\downarrow} \\
+ (U^{\prime}-J/2)   \sum_{i} n_{i,xz}n_{i,yz} -2J\sum_{i}{S}^z_{i,xz}{S}^z_{i,yz} \\
+ J^\prime \sum_{i}
(d^\dagger_{i,xz,\uparrow}d^\dagger_{i,xz,\downarrow}
d^{\phantom\dagger}_{i,yz,\downarrow}d^{\phantom\dagger}_{i,yz,\uparrow}+H.c.),
\end{aligned}
\end{equation}
where $d^{\dagger}_{i,\alpha,\sigma}$ 
creates an electron at site $i$, orbital $\alpha$ (either $xz$ or $yz$), and 
with spin projection $\sigma$. The number operator is $n_{i,\alpha,\sigma}$, 
$n_{i,\alpha}$=$\sum_{\sigma} n_{i,\alpha,\sigma}$, 
and ${S}^z_{i,\alpha}= (1/2)(n_{i,\alpha,\uparrow}- n_{i,\alpha,\downarrow})$.
$U$, $U^\prime$, $J$, and $J^\prime$ are the Kanamori parameters.
The usual constraints $U^{\prime}=U-2J$ and $J=J^{\prime}$ are assumed. In the Hund term
only the Ising portion is used because the expected magnetic order is collinear,
most materials have an easy-axis, and it is technically simpler for this
first study. The hopping parameters reproduce the Fermi surface 
of the undoped FeSC,\cite{raghu} but our conclusions could be realized 
as well in other materials with local tetragonal symmetry.
The crystal location of the Se, As, or P atoms, used by 
electrons to tunnel from Fe to Fe, justify 
that both nearest- (NN) and next-nearest-neighbor (NNN) hoppings are needed. 
The explicit hopping amplitudes are in previous publications.~\cite{raghu,moreo} 
The NN sites hoppings $t_1$ and $t_2$ are
only intraorbital. Along the plaquette diagonals, 
the intraorbital (interorbital) hopping is $t_3$ ($t_4$).
Their values are $t_1$=$1$, $t_2$=$-1.33$, and $t_3$=$t_4$=$-0.85$ and 
the bandwidth of this tight binding model is $W$=$12t_1$.\cite{raghu} Hereafter,
$t_1$ will be denoted by $t$, and it will be the energy unit. 
To convert to eV, the {\it ab-initio} derived bandwidth 
for the $d_{xz}$-$d_{yz}$ bands is $W \sim 1.8$ eV.\cite{fesc-bw-1,fesc-mass_renorm} 
The density is fixed to two electrons
per site ($n=2$).

Mean-field approximations have already been applied to two-orbital Hubbard models
at $T$=$0$,\cite{fesc-2orb-utk-1,fesc-2orb-utk-2} 
showing several phases with increasing $U/W$: a paramagnetic metal, 
a metal with ($\pi,0$) spin order, and an insulator also with ($\pi,0$) spin order. 
These previous $T$=$0$ studies and others~\cite{ironrefs,peter,dai,RMP} showed that 
$J/U\sim 0.15-0.30$ is relevant for FeSC, and 
we will fix $J/U=0.25$ in all the results below. Since 
FeSC materials vary substantially in their degree of electronic correlation,
results will be presented varying the ratio $U/W$. Our main focus are the temperature effects
since their influence on model Eq.~(1) are unknown.
Our study is performed in two dimensions, mainly on $32^2$ lattices.
The TCA traveling cluster~\cite{tca-2} is $6^2$. 
Via the parallelized version of TCA,\cite{tca-2} 
lattices as large as $48^2$ were reached. 
The mean-field approach was chosen to be the Hartree approximation that works
well for the half-filled one-orbital model.\cite{Jprime}

\begin{figure}[t]
\centering{
\includegraphics[width=9.5cm, height=6.66cm, clip=true]{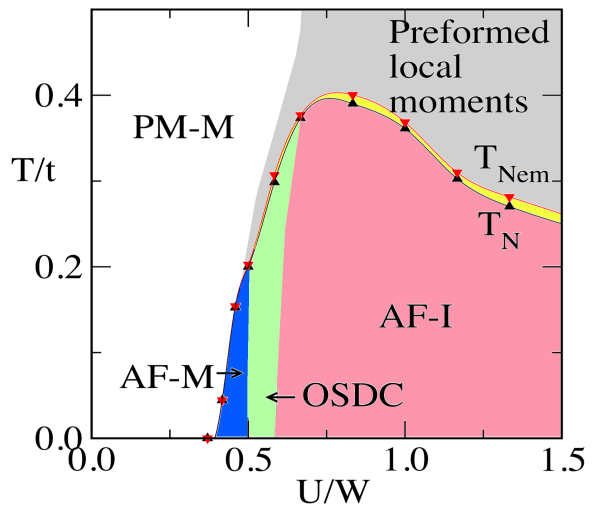}}
\caption{(color online) Phase diagram of the two-orbital Hubbard model Eq.(1) 
in the MC-MF approximation and 
with hoppings from previous literature,\cite{raghu} at $J/U=0.25$ and $n=2$.
Shown are results for a $32^2$ lattice. Besides the 
weak coupling paramagnetic metal (PM-M) and the intermediate/large $U/W$
region with ``preformed local moments'' (grey), other states were identified: 
(1) A $(\pi,0)$-spin-ordered metal, AF-M, where transport is anisotropic but 
metallic in both directions.
(2) A novel $(\pi,0)$-spin-ordered regime, the OSDC, with metallic (insulating) 
behavior along the $x$ ($y$) axis.
(3) A $(\pi,0)$-spin-ordered insulator, AF-I, with a full gap.
(4) A very narrow spin nematic regime above $T_N$.
The AF-M and OSCD states break the same symmetries and, thus,
no sharp distinction between them is expected but a rapid crossover. 
$T_N$ has a non-monotonic behavior, maximizing 
close to the OSDC/AF-I boundary. 
}
\label{rf-1}
\end{figure}

\section{Results}

Our main results are in the phase diagram of 
model Eq.(1) at $n=2$ (Fig.~\ref{rf-1}). It 
was constructed based on data for the two spin structure factors of relevance, $S(\pi,0)$ and 
$S(0,\pi)$, the spin nematic order parameter $\Psi_{Nem}$,\cite{nematicrefs}
their temperature derivatives, and the resistivity and density of states (DOS).
We also monitored local 
moment formation at intermediate temperature and large $U/W$.\cite{mfmc-rec-prb}
This phase diagram is surprisingly rich because it 
contains {\it three} regimes with $(\pi,0)$ long-range magnetic order 
[degenerate with $(0,\pi)$]. The insulator 
at large $U/W$ is induced by the robust $J$ that  produces
$S$=1 local spins interacting via a frustrated Heisenberg model
known to have $(\pi,0)$-$(0,\pi)$ magnetic order. The other two
states at intermediate couplings are more subtle. Intuitively, their 
magnetic order can be considered as arising from Fermi surface nesting effects. 
However, the presence of $two$, instead of one, regions is unexpected.
The first one, dubbed AF-M (antiferromagnetic metallic), 
is metallic in both directions, albeit anisotropic,
as observed in FeSC experiments.\cite{Chu} 
But the second region, the OSDC, is
metallic along the spin staggered direction but insulating along the
spin uniform direction,
leading to a surprising dimensional reduction at intermediate couplings.

\subsection{Magnetic phases} 

\begin{figure}[t]
\centering{
\includegraphics[width=9.cm, height=5.0cm, clip=true]{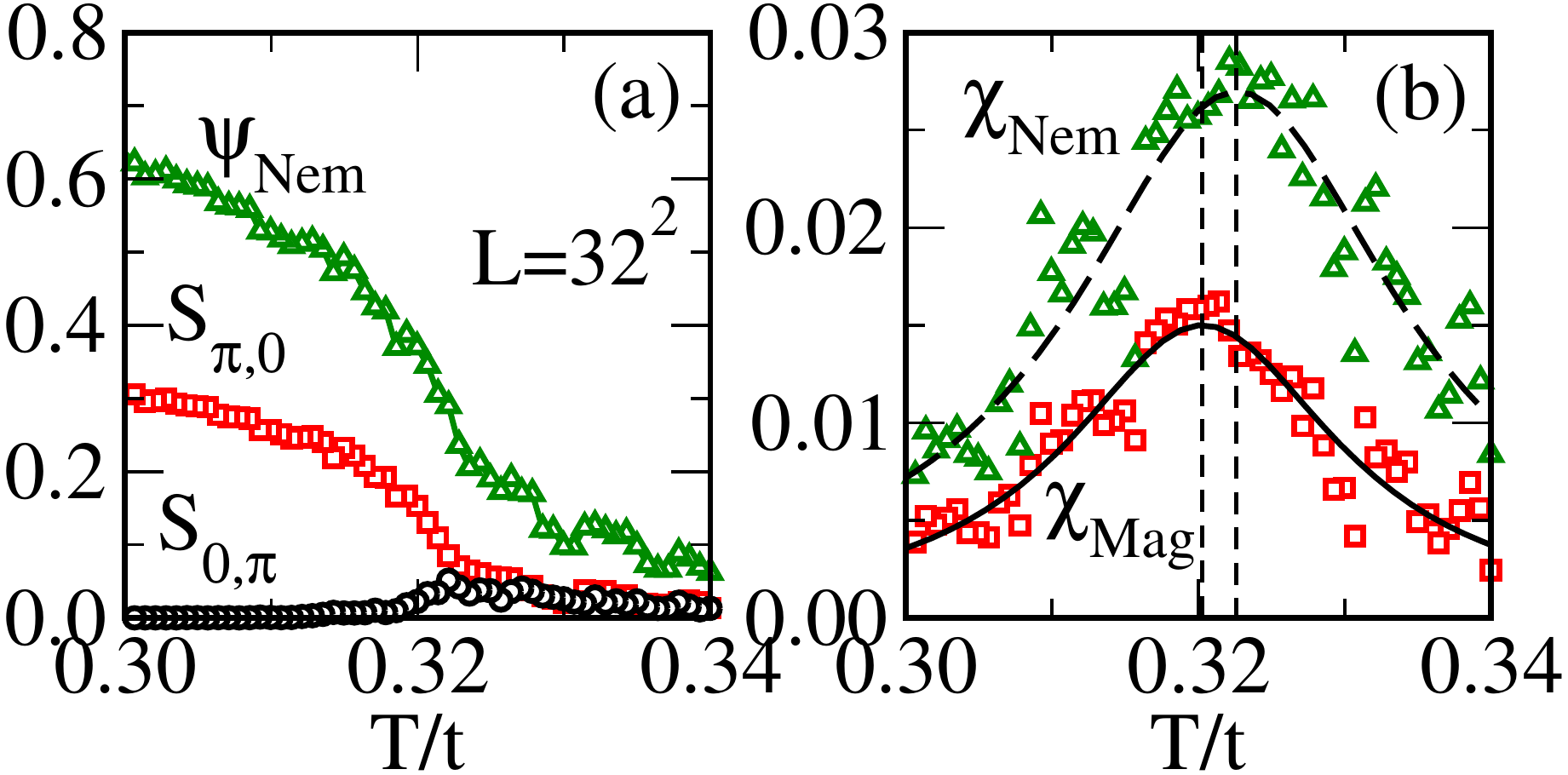}}
\caption{(color online) 
(a) Spin structure factors 
and nematic order parameter at $U/W=1.16$ and using a $32^2$ lattice (statistical errors bars
are the size of the points; error bars due to trapping in metastable states are
better represented by the fluctuations of the results with varying temperatures).
(b) Same as (a) but for the magnetic(for $S(\pi,0)$ case) and nematic susceptibilities. Lorentzian fits of the data 
are shown.
Vertical dashed lines (left to right) indicate  the magnetic and
nematic transition temperatures, from susceptibility maximization.
}
\label{rf-2}
\end{figure}

\begin{figure}[t]
\centering{
\includegraphics[width=8.5cm, height=11.00cm, clip=true]{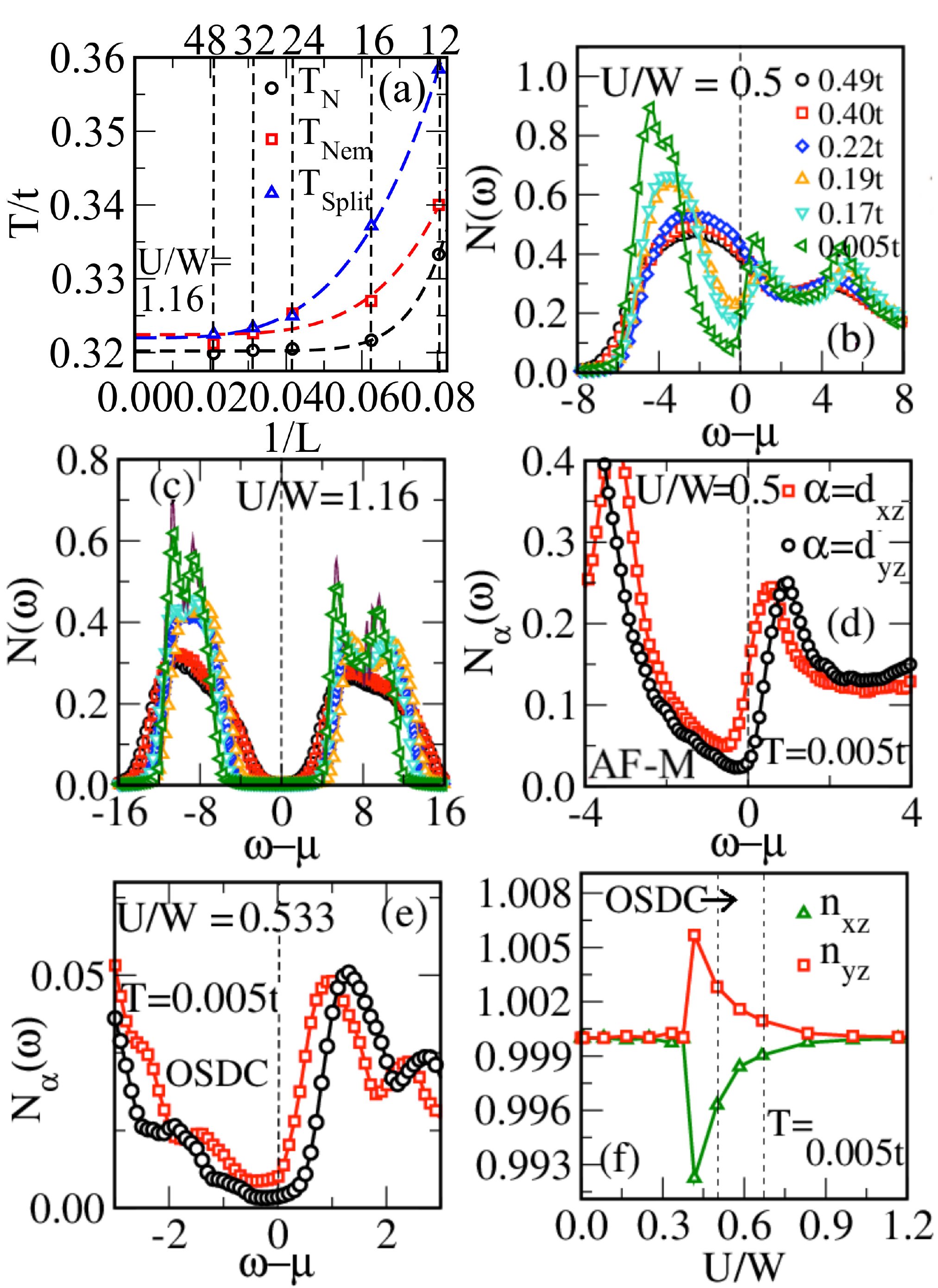}}
\caption{(color online) 
(a) Finite-size scaling analysis for $T_N$, $T_{Nem}$, and $T_{split}$ 
at $U/W=1.16$, using $L = 12, ..., 48$ ($L\times L$) lattices
plotted vs. $1/L$ (values at top). Data is fit with a   
scaling function $T_{ord}(L)=T_{ord}^{bulk}+b(1/L)^{(1/\nu)}$, where
$T_{ord}^{bulk}$, $b$, and $\nu$ are independent fitting parameters 
for the three data sets. From the fit 
we obtain a nematic temperature window of width approximately $0.002 t$.
(b) Density of states $N(\omega)$  at $U/W$=$0.5$ ($\mu$ denotes $E_F$). 
A pseudogap develops 
below $T_{Nem}$ that deepens with reducing temperature but never becomes a full gap.
On the other hand, in panel (c) at $U/W$=$1.16$ a clear gap develops 
upon cooling. (d) contains the orbital-resolved DOS at 
$T$=$0.005t$ and  $U/W$=$0.5$. The $C_4$ spontaneous symmetry breaking makes
the (nonzero) population of the two orbitals 
different at $E_F$.\cite{FSOO}
(e) Same as (d), but at $U/W$=$0.533$ in the OSDC regime where 
$N_{yz}(\omega = \mu) << N_{xz}(\omega = \mu)$.
(f) The orbital-resolved total 
occupation $n_{xz}$ and $n_{yz}$ shown at low $T$ vs.  $U/W$. Dashed lines
indicate the OSDC regime. Panels (b-f) were obtained using $32^2$ lattices.
}
\label{rf-4}
\end{figure}

A typical magnetic order parameter ($OP$) 
displays a negative curvature increasing temperature(Fig.~\ref{rf-2}a), with a diverging slope 
at the critical temperature $T_c$ in the bulk limit. Thus, for a finite system 
the temperature where the first derivative d$OP$/d$T$ is maximized, or where the
associated susceptibility maximizes, provides an
estimation of $T_c$ (here $T_N$ and $T_{Nem}$)(Fig.~\ref{rf-2}b). 
In addition, upon cooling $T_{Nem}$ can also be estimated
from the temperature $T_{split}$ where $S(\pi,0)$ and $S(0,\pi)$ split. While
for a finite system $T_{Nem}$ and $T_{split}$ may be different, 
they should merge in the bulk limit. 
Typical results are in Fig.~\ref{rf-2} for a $32^2$ lattice, at a fixed ratio $U/W$. 
Our MC statistics and lattice sizes are sufficient to observe a robust 
order parameter behavior for $S(\pi,0)$ and $\Psi_{Nem}$: 
nonzero at $T=0$ and decreasing with increasing
temperature with a negative curvature, with the exception of a small temperature window
where the curvature is positive due to size effects.
Although very close in temperature, systematically 
for all the couplings $U/W$ with long range spin order at $T=0$ 
and for all lattices, we find $T_{Nem}$ slightly larger than $T_N$ suggesting a small region of nematicity.
In general, we  also found that $T_{split}$ tends to be  larger than $T_{Nem}$.
The observed narrow window of nematicity in fact appears to survive a finite-size scaling analysis
(Fig.~\ref{rf-4}(a)), at least for the investigated coupling $U/W=1.16$ where information for lattices
as large as $48^2$ were gathered (since these simulations are very time consuming the
scaling analysis could be done only for one coupling).
Remarkably, within the error bars 
the bulk limit extrapolated $T_{Nem}$ and $T_{split}$ seem to lead to the same 
nematic critical temperature.
More specifically, our results unveil a small nematic window of $\sim 0.002 t$ at $U/W=1.16$.
The narrowness of this nematic regime is likely
exaggerated by the Z(2) nature of the Ising
approximation used here for the Hund term. Such a fragile nematic phase, 
reported here for the first time in a Hubbard model, is compatible
with previous studies using spin models,\cite{batista} 
spin-fermion models,\cite{prl2013} and with experiments,\cite{Chu,firstvssecond}
that have also reported very narrow nematic temperature windows.

\subsection{New State at Intermediate Coupling}

As explained before it is surprising that 
there are  three distinct regions below $T_N$: two metals (AF-M and OSDC) and one insulator (AF-I).
The distinction between the two metals and the insulator can be understood
via the density-of-states in Figs.~\ref{rf-4}(b,c). 
Panel (c) displays a canonical insulating behavior: 
in the temperature range shown, a pseudogap (PG) is observed 
in the local moments regime,\cite{comPG} 
transforming into a full gap when the magnetic order develops at $T_N$. 
This is the AF-I state (Fig.~\ref{rf-1}). In panel (b), upon cooling toward 
$T_N$ a pseudogap opens probably because of Fermi surface 
nesting effects. But even at low temperatures, 
and independently analyzing the zero-temperature Hartree equations, in both metallic regions the total DOS
has a finite weight at the Fermi energy $E_F$.

\begin{figure}[t]
\centering{
\includegraphics[width=8.5cm, height=10.0cm, clip=true]{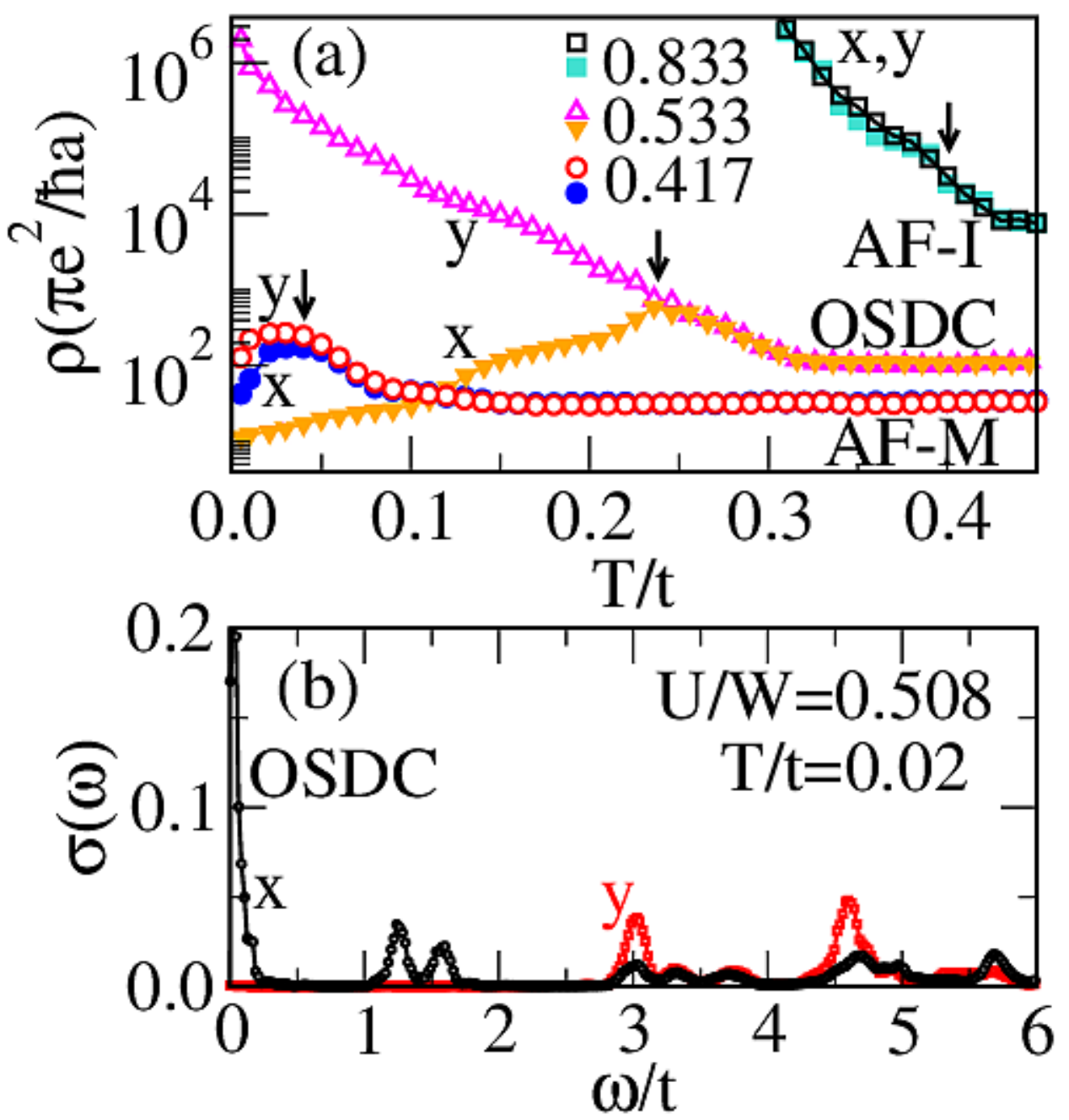}}
\caption{(color online) (a) Resistivity along the $x$ spin staggered 
(solid symbols) and $y$ spin uniform (open symbols)  
directions at several $U/W$'s, illustrating the transport properties of the 
AF-M, OSDC, and AF-I regions of Fig.~\ref{rf-1}. Arrows indicate $T_N$ for each case.
(b) Optical conductivity in the OSDC with 
electric fields along the $y$ and $x$ directions. 
In both panels a $16^2$ lattice is used.
}
\label{rf-5}
\end{figure}

What is then the difference between AF-M and OSDC? Their physical distinction
is illustrated in Fig.~\ref{rf-5}~(a) where the resistivity $\rho$ vs. $T$ is presented 
at three values of $U/W$, corresponding to the three low-temperature regions of Fig.~\ref{rf-1}.
$\rho$ is calculated from the optical conductivity $\sigma(\omega)$, integrating
in a narrow range near $\omega = 0$ and then inverting.\cite{sigma} 
At $U/W = 0.417$, $\rho$ in the 
$y$ spin uniform direction is larger than in the $x$ spin staggered
direction, as in previous calculations~\cite{prl2013,zhang,prl2012} and experiments.\cite{Chu}
This is understood from the orbital
resolved DOS of Fig.~\ref{rf-4}(d): in the magnetic $(\pi,0)$ state that breaks
rotational invariance, near $E_F$ the orbital $d_{yz}$, 
related to conduction in the
spin uniform direction, is more suppressed than $d_{xz}$, related 
to conduction in the spin staggered direction. But since 
both orbitals have a sizable DOS weight at the $E_F$,
both directions are metallic. At large 
$U/W = 1.0$, the DOS Fig.~\ref{rf-4}(c)
displays a sharp gap at low temperature, and both directions must be insulating.

Note that experiments affected primarily by the vicinity of $E_F$
may suggest strong orbital order, but
the $\omega$-integral of the orbital-resolved DOS, i.e. the orbital population $n_{xz}$ and $n_{yz}$,
is only different by 0.5\% (Fig.~\ref{rf-4}~(f)).\cite{FSOO} 
For completeness, we also monitored
$\Delta_{orb} = n_{xz}- n_{yz}$ vs. temperature. The results (not shown) 
indicate that upon cooling from high $T$,
$\Delta_{orb}$ remains smaller than 0.001 at $U/W$=$1.16$
until $T_{Nem}$ is reached, eventually converging, as $T$ is further reduced, to the results
derived from Fig.~\ref{rf-4}~(f). 
Within our MC accuracy we conclude that 
in our model there is no additional 
orbital-order critical temperature above $T_{Nem}$.

\begin{figure*}[t]
\centering{
\includegraphics[width=17.cm, height=13.cm, clip=true]{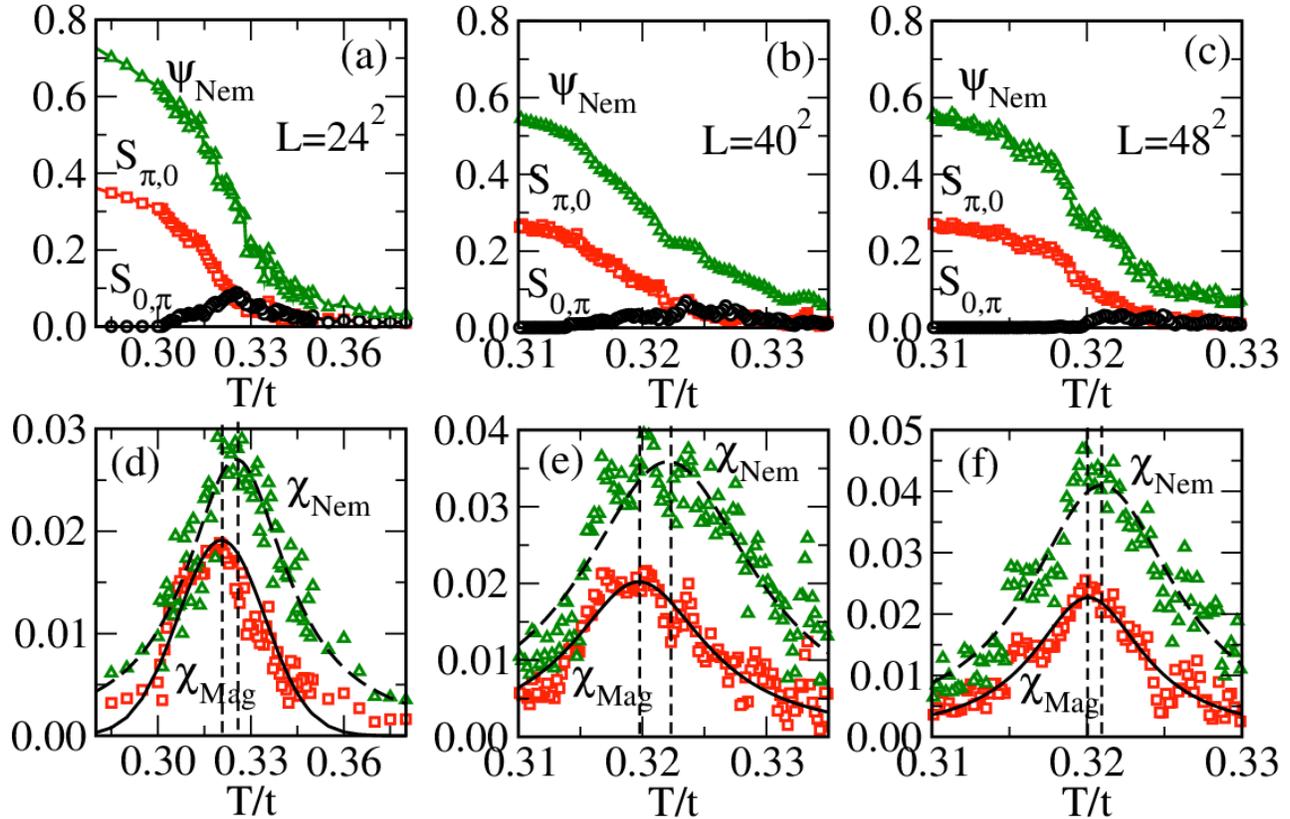}}
\caption{(color online) (a,b,c)
The $(\pi,0)$ and $(0,\pi)$ magnetic structure factors, and the 
nematic order parameter ($\Psi_{Nem}$) vs. temperature for 
different lattice sizes, at $U/W=1.16$. 
(d,e,f) Magnetic and nematic susceptibilities at $U/W = 1.16$ 
for the same lattice sizes as immediately above 
(i.e. (d) corresponds to (a), etc). The solid curves are Lorentzian fits.
The dashed lines indicate the positions of the maxima in the susceptibilities,
therefore indicating the estimated critical temperatures.
}
\label{srf-1}
\end{figure*}

The interpolation between small and large $U/W$ 
unveils a surprise: at intermediate couplings such as 
$U/W = 0.533$  the spin staggered direction remains metallic, but
the spin uniform direction becomes $insulating$ (Fig.~\ref{rf-5}~(a)).
Intuitively,
this is because the interpolation between the density-of-states of Fig.~\ref{rf-4}(d), where both orbitals
have nonzero weight at $E_F$, and (c), where both orbitals have negligible weight at $E_F$, is
not smooth. Instead, there is an intermediate coupling range where
the $d_{yz}$ weight at 
$E_F$ is almost negligible while that of $d_{xz}$ is still finite (Fig.~\ref{rf-4}(e)).
The $d_{yz}$ weight is very small but not zero, even at $T=0$, because of the broadening used for density of states.  Therefore, 
strictly speaking it cannot be used as a sharp order 
parameter: AF-M and OSDC are likely analytically connected
because they break the same symmetries.
However,
our study shows that the orbital population difference at $E_F$ in the OSDC is sufficiently
large to induce one-dimensional transport.
Results for the full $\sigma(\omega)$ (see Fig.~\ref{rf-5}~(b) and Appendix) 
show that 
at $\omega \sim 0$ only one orbital dominates.
Moreover, from $n_{xz}$ and $n_{yz}$ (Fig.~\ref{rf-4}~(f)) note that 
at large $U/W$ both orbital populations
converge to one since $J$ is large, 
while in the PM-M regime they are also one
by symmetry. However, in the OSDC region {\it both} 
$n_{xz}$ and $n_{yz}$ are different from one (and different among themselves 
because $C_4$ is spontaneously broken): the OSDC regime is 
not the same as an Orbital Selective Mott Phase~\cite{OSMP}
where one orbital has population exactly one. 
Another interesting observation is
that for both AF-M and OSDC there 
is an insulating region $d\rho/dT<0$ in both
directions immediately above $T_N$ due to the opening of the pseudogap 
in the local moments region and concomitant 
coexisting patches of ($\pi$,0) and (0,$\pi$) order,\cite{prl2012} 
in agreement with spin fermion studies~\cite{prl2012} 
and experiments.\cite{Chu}

\subsection{Other results used to construct the magnetic phase diagram}

The analysis presented in Fig.~\ref{rf-2} and Fig.~\ref{rf-4}(a) are just examples
of the vast computational study carried out 
at various values of $U/W$ and
used to construct the phase diagram of Fig.~\ref{rf-1}. For completeness, in Fig.~\ref{srf-1} 
 this substantial effort is illustrated further 
by providing data corresponding to other
lattice sizes at $U/W=1.16$. The
information gathered from these efforts for $T_N$, $T_{Nem}$, and $T_{split}$
were used for the finite-size scaling analysis of Fig.~\ref{rf-4}~(a).

\section{Conclusions} 

The phase diagram of a layered two-orbitals
Hubbard model was studied with emphasis on temperature effects.
We report a novel intermediate coupling $U$ 
region, called the OSDC, that is conducting in one direction via the
$d_{xz}$ orbitals, but insulating in the other because the associated 
$d_{yz}$ orbitals have nearly vanishing weight at $E_F$. 
Although our low temperature calculations  do not include
quantum fluctuations, the OSDC starts at relatively high temperatures 
$\sim T_N$ and for this reason our approach emphasizing thermal effects should be sufficient. Moreover,
we tested that using other hoppings,\cite{daghofer} the
OSDC is also found.
Experimentally, 
materials of the family of iron superconductors are the most likely to 
realize the OSDC (in fact, indications of OSDC are already 
present in Ref.~\onlinecite{Chu} for the case of doped 
materials), and a mixing 
in the chemical formula of As, associated with weak coupling, 
and Se, associated with strong coupling, may be needed.
But the OSDC could be realized in other layered materials where
a transition metal atom M is coordinated 
with four ligand atoms X, establishing 
MX$_4$ tetrahedral cages with near degenerate $d_{xz}$ and $d_{yz}$ orbitals,
and where a magnetic state that breaks lattice rotational invariance is
stabilized.

\section {Acknowledgments}

E. Dagotto thanks Guangkun Liu for useful conversations.
A. Mukherjee and N. Patel were  
supported  by  the  National  Science  Foundation  Grant
No. DMR-1404375. E. Dagotto and A. Moreo were supported by the
U.S. Department of Energy, Office of Basic Energy Science, 
Materials Science and Engineering Division.

\section{Appendix}

\subsection{Details of the Monte Carlo simulation and observables measured}

The lattices used all have periodic boundary conditions.
In our calculations, a total of 4000 MC system sweeps were typically performed: 2000 to thermalize 
the system, and the rest for calculating observables. A MC system sweep consists 
of sequentially visiting every lattice site  and updating the local auxiliary fields 
followed by the fermionic diagonalization or TCA procedure to accept/reject 
via the Metropolis algorithm. We start the simulation at high temperature with a 
random configuration of auxiliary fields and then slowly cool down to lower temperatures
to avoid being trapped in metastable states. 
The MC runs start at $T=1.0t$, which corresponds to about 1000~K and cool down to $T=0.005t$, where temperature steps as small as $\Delta T=0.0002t$ were used for the temperatures relevant to the magnetic and nematic transitions.
This slow process allows us to obtain reliable results independent of 
the initial conditions of the calculation. 

In our finite systems, there is no energy 
difference between the $(\pi,0)$ and $(0,\pi)$ magnetic states.
As a consequence, MC simulations that start at high temperature in a random state for the
auxiliary fields may end up in $(\pi,0)$ or $(0,\pi)$
with equal chance upon cooling. In practice, we simply discarded all cooling down
MC processes that led to a $(0,\pi)$ state at low temperatures. 

The antiferromagnetic order is studied via the spin structure factors,
\begin{equation}
S({\bold q}) = \frac{1}{L^4} \displaystyle\sum\limits_{i,j} e^{i {\bold q} \cdot ({\bold r}_{i}-{\bold r}_{j} )} 
\langle {{ S^z_i} { S^z_j}}\rangle. 
\end{equation} 
The two wavevectors of interest in FeSC are ${\bold q}=(\pi,0)$ and $(0,\pi)$.
The expectation value is generated by using the eigenvectors of the MC equilibrated configurations.
For the study of the nematic regime above $T_N$, we compute the nematic order parameter 
\begin{equation}
\Psi_{Nem}={{1}\over{2L^2}}\sum_{i,\pm}(S^z_{i} S^z_{{i}\pm {\hat{y}}}-S^z_{i} S^z_{{i}\pm {\hat{x}}}),
\end{equation}
where $\hat{x}$ and $\hat{y}$ are unit vectors connecting site ${i}$ 
with its nearest neighbors. The $\pm$ summation is 
over all nearest neighbors, and 
$\langle \Psi_{Nem}\rangle >0 $ in the $(\pi,0)$ magnetic phase. 
To better locate critical temperatures, 
we also calculate numerically the magnetic ($\chi_S$) and 
nematic ($\chi_{Nem}$) susceptibilities using the standard 
variance calculation.

In this work, we have also evaluated the orbital resolved density of states (DOS), 
$ N_{\alpha}(\omega) = \sum_{m} |\langle \xi_{m,\alpha}|\psi \rangle |^2\delta(\omega-\omega_m)$, 
where $\omega_m$ are the eigenvalues of the fermionic sector and the summation runs up to $2 L^2$, 
i.e. the total number of eigenvalues of a $L^2$ system with spin. 
$|\langle \xi_{m,\alpha}|\psi \rangle |^2$ is the weight of the $m^{th}$ eigenstate 
for orbital $\alpha$ in the state $|\psi\rangle$. $ N_{\alpha}(\omega)$ is calculated by implementing
the usual Lorentzian representation of the $\delta$ function. The broadening needed to 
obtain $N(\omega)$ from the Lorentzian is $\sim W/2L^2$, where $W$ 
is the fermionic bandwidth at $U=0$. Finally, $ N(\omega)$, the total DOS, is 
the sum of the different orbital densities of states. Numerically, e.g., 
for the 8$^2$ system
the broadening is about $0.09t$. Two hundred $N_{\alpha}(\omega)$ samples are obtained 
from the 2000 measurement system sweeps at every temperature. We discard 10 MC steps 
between measurements to reduce self-correlations in the data. The 200 $N_{\alpha}(\omega)$ 
samples are used to obtain the thermally averaged $\langle N_{\alpha}(\omega)\rangle_T$ 
at a fixed temperature. These are further averaged over data obtained from 10 independent 
runs with different random number seeds. A very similar procedure is followed to
calculate the optical conductivity, that involves matrix elements of the current operator.

To determine the crossover temperature between the weak coupling paramagnetic
state and the regime with preformed local moments 
above the magnetic order, we compute the specific heat, 
$C_v(U,J,T)=\frac{dE(U,J,T)}{dT}$, by numerically differentiating the average energy 
with respect to temperature, as well as the orbital resolved double occupation 
$\langle n_{\alpha,  \uparrow}n_{\alpha, \downarrow} \rangle$. 
The details of the procedure we followed is presented in our earlier work on the one-orbital Hubbard 
model,\cite{mfmc-rec-prb} but, briefly, $C_v(U,J,T)$ has a high-temperature peak 
that corresponds to local moment formation that can be tracked varying
$U$. In addition, the double occupation has to be 
below a cutoff for the system to have local moments.\cite{mfmc-rec-prb} This is 
of particular importance at small $U$ values because $C_v(U,J,T)$ can have a 
considerable contribution from the electronic delocalization and in this regime it cannot
be used to track local moment formation.

\begin{figure}[t]
\centering{
\includegraphics[width=9cm, height=7.5cm, clip=true]{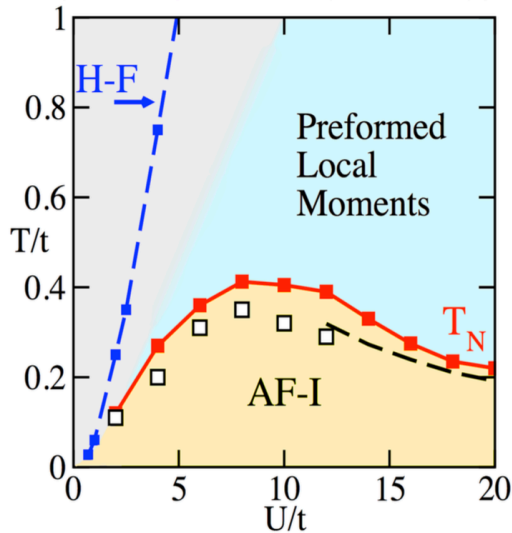}}
\caption{(color online) The Hubbard $U/t$ vs. temperature $T/t$ phase 
diagram corresponding to the one-orbital Hubbard model in three dimensions 
using the Hartree approximation in the MC-MF technique. $T_N$ denotes the
N\'eel critical temperature where antiferromagnetism with 
$(\pi, \pi,\pi )$ staggered magnetic order develops (region denoted as AF-I)
according to measurements of the spin structure factor and the spin-spin correlation functions.
The MC-MF results (red filled squares) are compared
against data obtained using the Determinant Quantum Monte Carlo [white squares, reproduced from 
R. Staudt \textit{et. al}, Eur. Phys. J. \textbf{B} 17, 411 (2000)]. The lattice
size used in the MC-MF method is $4^3$. The expected $4t^2/U$ behavior at large $U/t$ 
is indicated by the black dashed line.
The more crude mean-field Hartree Fock
approximation results are denoted by ``H-F'' (blue dashed line): it incorrectly
predicts the growth of the critical temperature with increasing $U/t$.
In the light blue region measurements of the spin square operator and the double
occupancy indicate the presence of a local moment.\cite{mfmc-rec-prb}
}
\label{srf-4}
\end{figure}

\subsection{Test of the Technique in the Hartree Approximation}

The many-body technique used in this publication was already
introduced and tested in previous efforts.\cite{mfmc-rec-prb} However,
since an easy axis is present in most materials and considering that
empirically the Monte Carlo convergence is improved under such 
circumstances (colloquially, Ising is easier than Heisenberg), in this
project it was decided to use the Ising approximation in the Hund term.
In the language of the one-orbital Hubbard model that corresponds to using
the Hartree approximation instead of
the Hartree-Fock approximation employed before.
This requires, then, a test of the Hartree assumption for 
the three-dimensional one-orbital standard 
Hubbard model at half-filling. There is no need to provide explicitly the
Hamiltonian for such well  known model, thus we move immediately to discuss
the results, which are provided in Fig.~\ref{srf-4}.

The results shown in Fig.~\ref{srf-4} are encouraging. The critical temperatures found
with the MC-MF technique capture the ``up and down'' behavior of $T_N$ with increasing
$U/t$ and they converge close to the expected  scaling at large $U/t$. Moreover,
in the entire range of $U/t$ investigated the MC-MF results are close to those of
quantum Monte Carlo (with the largest discrepancy being about 20\%).
The successful test presented in Fig.~\ref{srf-4} suggests that the MC-MF technique captures
the essence of the one-orbital problem, not only qualitatively but also quantitatively.
This gives us confidence that the results for two orbitals in the main text, 
that have not been studied before in the literature, are reliable. 

\subsection{Optical Conductivity}

\begin{figure}[t]
\centering{
\includegraphics[width=9.0cm, height=11.6cm, clip=true]{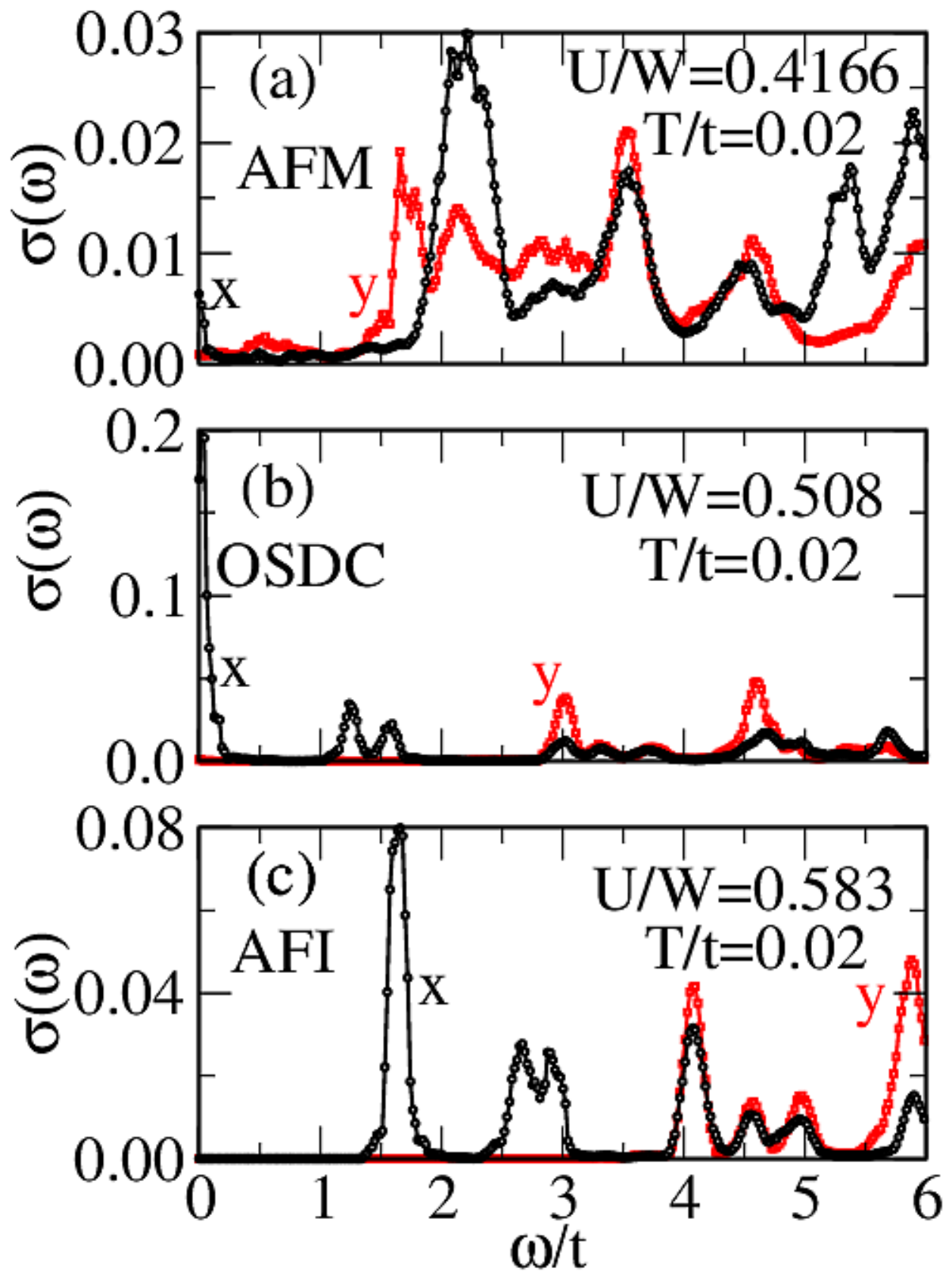}}
\caption{(color online) Optical conductivity of the 2D two-orbital model used in this study
at the temperatures and couplings $U/W$ indicated. Panel (a) is in the AF-M regime;
panel (b) in the OSDC regime; and panel (c) in the AF-I state.
All data are using a $16^2$ system. 
}
\label{srf-3}
\end{figure}

To illustrate the physics of the three different states reported here,
the optical conductivity was calculated. The results are in Fig.~\ref{srf-3}.
Panel (a) is in the AF-M regime: while the $(\pi,0)$ magnetic order breaks
the symmetry between the $x$ and $y$ directions, the difference is not dramatic
and leads to both directions being metallic.
Panel (c) corresponds to the Mott insulating regime: here a gap is present both
when the electric field points along the $x$ and $y$ directions. 
The first excitations occur at the scale of the Hund coupling, 
along the $x$ spin staggered
direction. The most novel result 
is shown in panel (b), already presented in the
main text and reproduced here for the benefit 
of the readers, corresponding to the new
OSDC region: 
here at $\omega/t \sim 0$ there is a finite weight in the $x$ direction 
but negligible weight in the $y$ direction, compatible with the calculation
of the resistivity shown in the main text.

Some subtle technical details are worth discussing, particularly with
regards to the window in $\omega$, around zero, used to define the resistivity.
Consider a $L^2$ lattice. Its associated mean level spacing 
is roughly estimated as $s=W/(4L^2)$ where $W$ is the bandwidth. In practice
the individual $\delta$-functions broadening used in the conductivity 
calculation is $s$ times a factor, which it has been chosen to be 4 in our
calculations. As example, for a $20^2$ system $s=12/(4 \times 20^2)=0.0075$, and 
the broadening used is then 0.03. The integration range is decided as follows: 
the smallest frequency (the starting $\omega$) is chosen to be at least 
one order of 
magnitude smaller than $s$. The frequency increment is chosen to be $s/10$. 
The data of these 10 frequency points is used for the integration and the 
outcome is ascribed as the average conductivity for the frequency 
value at the lower end of the interval.
In the example given above, then the frequency step is 0.003. 
The integration is performed over 10 frequency steps. 
The calculation is started at an initial frequency of $0.0001t$.

\subsection{Mean field formalism }

For completeness, here details of the mean-field approximation are provided.
In general, the multiorbital Hubbard model shown in the main text can be written as follows:

\begin{equation}
\begin{aligned}
H=H_o+H_{int}=
\sum_{\langle i,j \rangle,\alpha,\beta,\sigma} T^{i,j}_{\alpha,\beta} d^\dagger_{i,\alpha,\sigma}
d^{\phantom\dagger}_{j,\beta,\sigma} + \\
\sum_{i,\sigma,\sigma^\prime}\sum_{\alpha,\alpha^\prime,\beta,\beta^\prime}
U_{\sigma,\sigma^\prime}({\alpha,\alpha^\prime,\beta,\beta^\prime}) 
d^{\dagger}_{i,\alpha,\sigma}d^{\dagger}_{i,\alpha^\prime,\sigma^\prime}
d^{\phantom\dagger}_{i,\beta^\prime,\sigma^\prime}
d^{\phantom\dagger}_{i,\beta,\sigma}. 
\end{aligned}
\end{equation}
$H_o$ ($H_{int}$)  in the kinetic (interaction) term. $d^{\dagger}_{i,\beta,\sigma}$ 
creates an electron at site $i^{th}$, orbital $\beta$, and 
with spin projection $\sigma$. The onsite interaction is:

\begin{eqnarray}
U_{\sigma,\sigma^\prime}({\alpha,\alpha^\prime,\beta,\beta^\prime}) &=& 
\frac{U}{2} \delta_{-\sigma,\sigma^\prime}\delta_{\alpha,\alpha^\prime}\delta_{\alpha\beta}\delta_{\alpha\beta^\prime}  \\\nonumber
&+&\frac{U^\prime}{2}(1-\delta_{\alpha\alpha^\prime}) \delta_{\alpha\beta}\delta_{\alpha^\prime\beta^\prime} \\\nonumber
&+&\frac{J}{2}(1-\delta_{\alpha\alpha^\prime})\delta_{\alpha\beta^\prime}\delta_{\alpha^\prime\beta} \\\nonumber
&+& 
\frac{J^\prime}{2}\delta_{\alpha\alpha^\prime}\delta_{\beta\beta^\prime}(1-\delta_{\sigma\sigma^\prime})(1-\delta_{\alpha\beta}).
\end{eqnarray}

To derive the MF Hamiltonian we follow the treatment 
in {\it {Quantum Theory of Finite Systems}}, 
by Blaizot, J.-P. \&  Ripka, G., The MIT Press (1985).
The advantage of this approach is that one can derive a single general expression for the mean field parameters, for any number of orbitals. This expression can be easily coded in, thus avoiding the need to derive all possible mean field decouplings by hand. For this purpose we introduce the following notation:
\begin{eqnarray}
\rho_{i,j,\alpha,\beta} & = & \langle\alpha|\rho_{i,j}|\beta\rangle=\langle d_{j,\beta}^{\dagger}d_{i,\alpha}\rangle,
\end{eqnarray}
where $\rho_{i,j,\alpha,\beta}$ are elements of the single particle density matrix. We now make the Hartree-Fock approximation assumption: the state of the system can be represented by a single Slater determinant,  $|\Psi\rangle$. By using the Wick's theorem, we can then write down the expectation value of $H$ in $|\Psi\rangle$, denoted by $E[\rho]$,  as: 
\begin{equation}
\begin{aligned}
E[\rho]= \sum_{\langle i,j \rangle,\alpha,\beta,\sigma} T^{i,j}_{\alpha,\beta}\langle\beta,\sigma|\rho_{i,j}|\alpha,\sigma\rangle +\\ 
\sum_{i,\sigma,\sigma^\prime}\sum_{\alpha,\alpha^\prime,\beta,\beta^\prime}U_{\sigma,\sigma^{\prime}}(\alpha,\alpha^{\prime},\beta,\beta^{\prime})\times\\ [\langle\beta,\sigma|\rho_{i,i}|\alpha,\sigma\rangle\langle\beta^{\prime},\sigma^{\prime}|\rho_{i,i}|\alpha^{\prime},\sigma^{\prime}\rangle\\
-
\langle\beta^{\prime},\sigma^{\prime}|\rho_{i,i}|\alpha,\sigma\rangle\langle\beta,\sigma|\rho_{i,i}|\alpha^{\prime},\sigma^{\prime}\rangle].
\end{aligned}
\end{equation}

We now take the derivative of $E[\rho]$ with respect to a generic density matrix element, $\langle\bar{\alpha},\bar{\sigma}|\rho_{i,i}|\bar{\beta},\bar{\sigma^{\prime}}\rangle$, to get an explicit formula for Hartree-Fock mean field parameters, $\langle\bar{\beta},\bar{\sigma}^{\prime}|h|\bar{\alpha,}\bar{\sigma}\rangle$. Running over all values of orbitals and spin in $\langle\bar{\alpha},\bar{\sigma}|\rho_{i,i}|\bar{\beta},\bar{\sigma^{\prime}}\rangle$, for taking the derivatives, generates all possible mean field decouplings in the Hartree-Fock channel. Thus, the general formula is as follows:

\begin{equation} 
\begin{aligned}
\langle\bar{\beta},\bar{\sigma}^{\prime}|h|\bar{\alpha,}\bar{\sigma}\rangle=\sum_{\alpha',\beta'}(\sum_{\sigma'}~~~~~~~~~~~~~~~~~~~~~~~~~~~~~~\\
[U_{\bar{\sigma}\sigma'}(\bar{\beta},\alpha',\bar{\alpha,}\beta')\delta_{\bar{\sigma}\bar{\sigma}'}+U_{\sigma'\bar{\sigma}}(\alpha',\bar{\beta},\beta',\bar{\alpha}) \delta_{\bar{\sigma}\bar{\sigma}'}]\times \\ \langle\beta^{\prime},\sigma^{\prime}|\rho_{i,i}|\alpha^{\prime},\sigma^{\prime}\rangle\\
-[U_{\bar{\sigma}'\bar{\sigma}}(\bar{\beta},\alpha',\beta',\bar{\alpha})+U_{\bar{\sigma}\bar{\sigma}'}(\alpha',\bar{\beta},\bar{\alpha,}\beta')]\times \\ \langle\beta^{\prime},\bar{\sigma}^{\prime}|\rho_{i,i}|\alpha^{\prime},\bar{\sigma}\rangle).
\end{aligned}
\end{equation}
Finally, the full Hartree-Fock Hamiltonian is given by:
\begin{equation} 
\begin{aligned}
h_{MF}=\sum_{\langle i,j \rangle,\alpha,\beta,\sigma} T^{i,j}_{\alpha,\beta} d^\dagger_{i,\alpha,\sigma}
d^{\phantom\dagger}_{j,\beta,\sigma}\\ +\sum_{i,\bar{\alpha},\bar{\sigma}}\sum_{\bar{\beta}\bar{\sigma'}}\langle\bar{\beta},\bar{\sigma}^{\prime}|h|\bar{\alpha,}\bar{\sigma}\rangle d^\dagger_{i,\bar{\alpha},\bar{\sigma}}
d^{\phantom\dagger}_{i,\bar{\beta}\bar{\sigma'}} 
\\ 
-\sum_{i,\sigma,\sigma^\prime}\sum_{\alpha,\alpha^\prime,\beta,\beta^\prime}U_{\sigma,\sigma^{\prime}}(\alpha,\alpha^{\prime},\beta,\beta^{\prime})\times\\ ( \langle\beta,\sigma|\rho_{i,i}|\alpha,\sigma\rangle\langle\beta^{\prime},\sigma^{\prime}|\rho_{i,i}|\alpha^{\prime},\sigma^{\prime}\rangle\\
-\langle\beta^{\prime},\sigma^{\prime}|\rho_{i,i}|\alpha,\sigma\rangle\langle\beta,\sigma|\rho_{i,i}|\alpha^{\prime},\sigma^{\prime}\rangle).
\end{aligned}
\end{equation}
\\


\begin{thebibliography}{99}
\bibitem{fesc-mass_renorm}
Note that the actual bandwidth is likely to be smaller as 
a result of a significant mass renormalization 
in the undoped FeSC materials.
\bibitem{Jprime} In the Hartree approximation the $J^\prime$ term
does not contribute, but this term is only 
considered of marginal importance in Hubbard models.
\bibitem{firstvssecond}
General arguments~\cite{nematicrefs} suggest that either $T_N = T_{Nem}$ in a 
first-order transition, or $T_N < T_{Nem}$ with both transitions 
being second order. Since we do not observe indications of
metastabilities in the MC time evolution, or in histograms of observables, 
our results are compatible with the existence of a nematic regime.
\end{thebibliography}
\end{document}